\documentclass[lettersize,journal]{IEEEtran}

\usepackage{amsmath,amsfonts}
\usepackage{algorithmic}
\usepackage{algorithm}
\usepackage{array}
\usepackage{textcomp}
\usepackage{stfloats}
\usepackage{url}
\usepackage{verbatim}
\usepackage{graphicx}
\usepackage{cite}
\usepackage{booktabs}
\usepackage{multirow}
\usepackage{tabularx}
\usepackage{pdfcomment}

\usepackage{orcidlink}

\usepackage[
    caption=false,
    font=normalsize,
    labelfont=sf,
    textfont=sf
]{subfig}

\begin{document}

\title{Gen-TAS: A Generative AI-Aided Hardware-Software Task Allocation Framework for FPGA-GPP Heterogeneous Systems}

\author{
    Mary Kong\textsuperscript{\dag} \textsuperscript{\orcidlink{0009-0006-9082-6413}}, \IEEEmembership{Student Member, IEEE},
    Yuqin Zhao\textsuperscript{\dag} \textsuperscript{\orcidlink{0000-0001-7943-1433}},
    Semih Vazgecen \textsuperscript{\orcidlink{0009-0003-8868-3423}}, \\
    Cristian Sestito \textsuperscript{\orcidlink{0000-0002-7731-0002}}, \IEEEmembership{Member, IEEE}, 
    and Themis Prodromakis \textsuperscript{\orcidlink{0000-0002-6267-6909}}, \IEEEmembership{Senior Member, IEEE} \\

\thanks{
\textit{\textsuperscript{\dag}Yuqin Zhao and Mary Kong contributed equally to this work. Corresponding author: Yuqin Zhao (email:yzhao8@ed.ac.uk)}


This work was supported by the UKRI-EPSRC APRIL AI Hub (EP/Y029763/1) and the 2026 Google DeepMind Research Ready Scheme (GDMRR-2526-2-107), supported by Google DeepMind, The Hg Foundation, DSIT, and the Royal Academy of Engineering.

The authors are with the Centre for Electronics Frontiers, Institute for Integrated Micro and Nano Systems, School of Engineering, University of Edinburgh, UK.

}

}

\maketitle

\begin{abstract}

FPGA-GPP heterogeneous systems combine software flexibility with the performance and energy efficiency of reconfigurable hardware. However, determining which application tasks should execute on the GPP or FPGA requires extensive expertise and design-space exploration, particularly when user objectives vary across latency, communication, resource utilisation, and power. This paper proposes Gen-TAS, a knowledge-grounded LLM framework for user-specific FPGA-GPP task allocation. By combining task-graph analysis with RAG, Gen-TAS grounds LLM reasoning in historical implementation knowledge and generates multiple explainable strategies tailored to the specified objectives. Human-in-the-loop selection and a deterministic backend connect LLM-generated decisions to reproducible FPGA SoC implementations. Experiments on CNN and SDR workloads across multiple LLMs demonstrate stable, requirement-driven allocation. Under latency-oriented objectives, implementations following the selected strategies achieve speedups of up to 2.45$\times$ and 92.53$\times$, respectively, relative to the corresponding all-GPP baselines while other objectives select strategies that trade some acceleration performance for FPGA-GPP communication, resource utilisation, or FPGA power.

\end{abstract}

\section{Introduction}

\IEEEPARstart{H}{eterogeneous} field-programmable gate array (FPGA)-general-purpose processor (GPP) systems combine the programmability of software execution with the performance and energy efficiency of reconfigurable hardware. This complementary architecture is well suited to embedded acceleration, allowing control-oriented tasks to execute on the GPP while latency-, throughput-, and power-critical functions are mapped to FPGA-based custom datapaths \cite{ref12}. FPGA acceleration has demonstrated substantial benefits for Deep Neural Network (DNN) and Convolutional Neural Network (CNN) \cite{ref1,ref3}, sparse and resource-aware neural architectures \cite{ref5}, and software-defined radio (SDR) workloads \cite{CAMC-FPGA}. However, identifying an effective allocation of application tasks between the GPP and FPGA remains a significant design challenge.

Hardware-software partitioning directly affects latency, resource utilisation, memory access, communication overhead, interface design, and system-level performance. Prior codesign studies show that mapping and scheduling decisions strongly influence the quality of heterogeneous implementations \cite{ref13,ref14}. In practice, designers must analyse source code, identify computation stages and dependencies, estimate hardware and software costs, explore different mappings, and repeatedly interact with FPGA implementation tools \cite{ref15}. This process becomes more difficult when user requirements vary, since latency-oriented, resource-constrained, and power-aware designs may require different allocation strategies.
\begin{table}[!t]
\centering
\caption{Comparison with FPGA and FPGA-GPP partitioning methods.}
\label{tab:tool_comparison}
\vspace{-2mm}

\begingroup
\fontsize{7pt}{7.4pt}\selectfont
\setlength{\tabcolsep}{1.5pt}
\renewcommand{\arraystretch}{0.78}

\setlength{\aboverulesep}{0.15ex}
\setlength{\belowrulesep}{0.15ex}
\setlength{\cmidrulesep}{0ex}
\setlength{\cmidrulewidth}{0.3pt}

\begin{tabularx}{\columnwidth}{
    >{\centering\arraybackslash}p{0.17\columnwidth}
    >{\centering\arraybackslash}p{0.17\columnwidth}
    >{\centering\arraybackslash}p{0.18\columnwidth}
    >{\centering\arraybackslash}X
}
\toprule
\textbf{Method}
& \textbf{Platform / Scope}
& \textbf{AI / Knowledge}
& \textbf{Limitation or differentiator} \\
\midrule

EcoSys \cite{ref19}
& FPGA-GPP/ DNN video analytics
& None/ profiling and analytical models
& Application-specific and profiling-based; limited requirement interpretation \\

\midrule

StreamBlocks \cite{ref20}
& FPGA-GPP/ streaming tasks
& None/ execution profiles
& Profile-guided task parallelism; restricted to streaming applications \\

\midrule

CariPy \cite{CariPy}
& FPGA/ accelerator generation
& Generative AI/ knowledge base
& Hardware-generation focused; no HW/SW boundary reasoning or heterogeneous deployment \\

\midrule

SECDA-DSE \cite{SECDA-DSE}
& FPGA/ HLS accelerator DSE
& Generative AI/ KB and RAG
& Accelerator-template DSE; no general HW/SW task allocation \\

\midrule

\textbf{Gen-TAS (This Work)}
& FPGA-GPP/ function-level allocation
& Agentic AI/ KB and RAG
& Requirement-driven; explainable allocation and deterministic implementation \\[-0.4mm]

\bottomrule
\end{tabularx}

\endgroup
\vspace{-1.5mm}
\end{table}

Existing approaches have addressed related parts of this problem, but not the complete requirement-driven FPGA-GPP allocation flow. FPGA-GPP works such as EcoSys and StreamBlocks explore hardware-software partitioning using profiling, analytical models, or task-parallel execution information \cite{ref19,ref20}. FPGA accelerator frameworks, including Agamotto and SECDA-DSE, focus on accelerator optimisation or design-space exploration for specific FPGA-oriented implementations \cite{ref2, SECDA-DSE}. Recent LLM-assisted EDA methods further show the potential of generative AI for hardware design generation, EDA automation, and FPGA accelerator exploration \cite{CariPy,Lamda,ref17,ref18,ref22}. 
Nevertheless, Table~\ref{tab:tool_comparison} shows that existing state-of-the-art methods remain application-specific, FPGA-centric, or focused on design generation rather than general FPGA-GPP task allocation. 

\begin{figure*}[!t]
    \centering
    \includegraphics[
        width=1\textwidth,
        trim={1.4cm 8.1cm 1.2cm 4.7cm},
        clip
    ]{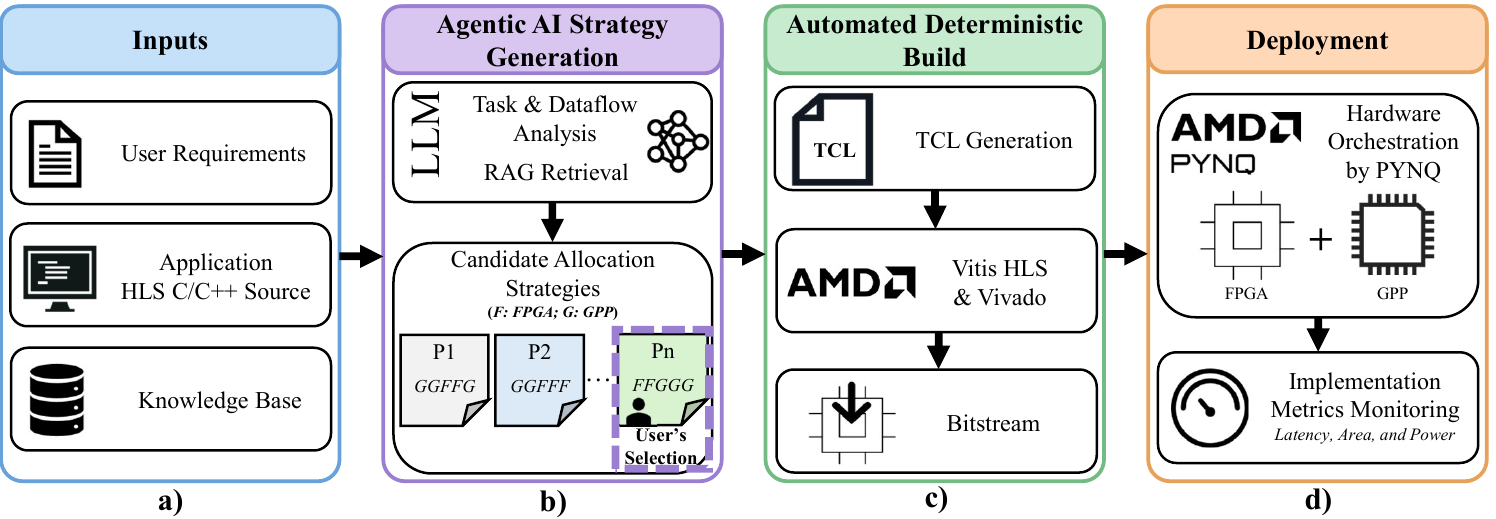}
    
    \caption{Gen-TAS workflow: (a) design-context construction from user requirements, HLS code, and implementation knowledge; (b) agentic AI with task/dataflow analysis and RAG for FPGA--GPP partition generation and user selection; (c) deterministic Vitis HLS/Vivado implementation of the selected strategy; and (d) PYNQ-based execution and measurement of latency, resources, and power.}
    \label{fig:framework}
\end{figure*}

In this work, we propose Gen-TAS, an LLM-assisted framework for requirement-driven, function-level task allocation across FPGA and GPP execution domains. As shown in Table~\ref{tab:tool_comparison}, Gen-TAS advances beyond prior methods by integrating source-code task-graph extraction, knowledge-base (KB) retrieval, retrieval-augmented generation (RAG) -grounded allocation reasoning, explainable strategy generation, human-in-the-loop selection, and deterministic Vivado/Vitis HLS and PYNQ implementation. Evaluation across two workloads, four design objectives, and three LLMs demonstrates requirements-oriented, stable, and reproducible allocation generation, achieving speedups of up to 2.45$\times$ for ImageProc and 92.53$\times$ for CAMC. Source code and supplementary materials are available at \url{https://github.com/aprilaihub/Gen-TAS}
\begin{figure}[!t]
    \centering
    \includegraphics[
        width=1\columnwidth,
        trim={0.6cm 2.5cm 0.2cm 2.5cm},
        clip
    ]{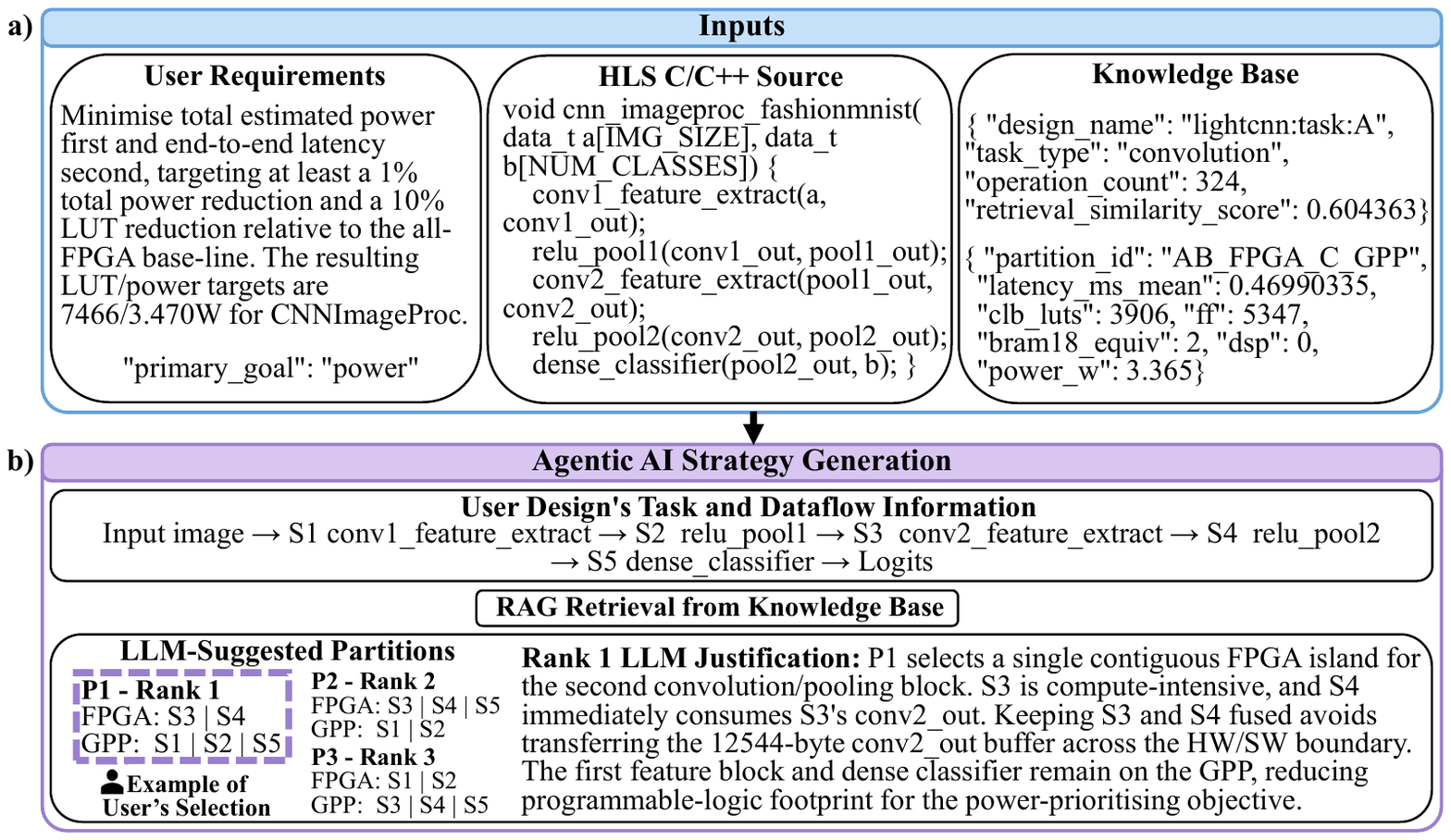}

    \caption{Requirement- and retrieval-driven FPGA-GPP task allocation in Gen-TAS: (a) design-context construction from natural-language requirements, HLS C/C++ code, and retrieved implementation knowledge; and (b) agentic AI combining task/dataflow analysis, RAG, and an LLM to generate, justify, and rank candidate partitions for user selection before deterministic implementation and hardware validation.}
    \label{fig:partition}
\end{figure}

\section{Gen-TAS Methodology}
\label{sec:methodology}

\subsection{System Model and Overall Framework}

Gen-TAS performs function-level task allocation for heterogeneous FPGA-GPP systems, where each computational function or processing stage identified in the high-level synthesis (HLS) C/C++ source code is assigned to either GPP execution or FPGA acceleration. As shown in Fig.~\ref{fig:framework}, the user provides the application source code and a natural-language design objective, while Gen-TAS accesses an internal knowledge base (KB) containing previously characterised implementations to support allocation decisions. The application is represented as an ordered task graph $G=(\mathcal{T},\mathcal{E})$, where $\mathcal{T}=\{T_1,\ldots,T_M\}$ contains the $M$ partitionable sub-functions and $\mathcal{E}$ their data dependencies. Here, $T_i$ denotes the $i$-th task. A candidate partition $P_j$ represents one possible hardware-software allocation, assigning each application task $T_i$ to either the GPP or FPGA, i.e., $P_j(T_i)\in\{\mathrm{GPP},\mathrm{FPGA}\}$. For hardware generation, the current backend maps a contiguous group of FPGA-assigned tasks to a single hardware region, while the remaining tasks execute on the GPP. Design objectives may target latency, power, FPGA resources, communication overhead, or a balance of these metrics.

The Gen-TAS workflow comprises source analysis, knowledge retrieval, allocation generation, and deterministic implementation, followed by on-board deployment and evaluation. Source analysis constructs the task-level design descriptors, which are used to retrieve relevant implementation evidence from the KB. The task information, retrieved evidence, platform constraints, and design objectives are supplied to a LLM to generate and rank candidate FPGA-GPP allocations. Valid candidates are presented to the designer for selection, after which a deterministic backend generates the corresponding hardware-software implementation. The LLM therefore suggests allocations, while implementation is deterministic.

\subsection{Task Analysis and Allocation Generation}
\label{subsec:partitioning}

Fig.~\ref{fig:partition} shows more details on the allocation procedure. Gen-TAS identifies the sub-functions called by the application top function as partitionable tasks. Task dependencies are derived from the call sequence and inter-task data flow. For each $T_i$, the extracted descriptor records operation type, input and output dimensions, numerical precision, buffer sizes. Communication volume is derived from buffer size and data precision. A hardware-software communication boundary occurs when two dependent tasks are assigned to different execution domains, requiring intermediate data transfer between the processing system and programmable logic (PS-PL).

The KB contains task characterisations and measured partition-level implementations. Records include computation and communication characteristics, FPGA-GPP placement, latency, look-up table (LUT), flip-flop (FF), digital signal processor (DSP), and block random-access memory (BRAM) utilisation, timing, power, communication boundaries, and, where available, measured transfer latency. Resource and timing data are obtained from Vitis HLS and Vivado reports, while execution and transfer latencies are obtained from on-board measurements. These records support allocation decisions through prior implementation evidence, while target-workload performance is determined through implementation and measurement.

During knowledge retrieval, Gen-TAS forms a query from the user request, parsed requirements, and extracted task characteristics. All KB records are scored against this query using token-based cosine similarity, and the 12 highest-scoring records are provided to the LLM as context. No minimum similarity threshold is imposed. When useful historical evidence is unavailable, allocation relies on the source-derived task characteristics and platform constraints.

The LLM prompt combines the task graph, task descriptors, retrieved records, platform constraints, and design objective. The LLM returns a configurable number $N$ of ranked candidates $\mathcal{P}={P_1,\ldots,P_N}$, where $\mathcal{P}$ denotes the candidate set, $P_j$ its $j$-th partition, and $N$ is set to three by default. Each candidate specifies the FPGA and GPP task sets, resulting communication boundaries, and a rationale based on the supplied source and KB evidence. Both acceleration potential and boundary communication are considered, since FPGA offloading can increase end-to-end latency when transfer and execution overheads exceed the computational benefit.

\subsection{Partition Validation and Deterministic Implementation}
\label{subsec:implementation}

Before user selection, the generated partitions are deterministically validated: all task identifiers must be valid, every task must be assigned exactly once, the FPGA and GPP sets must be disjoint, and FPGA-assigned tasks must form a contiguous group in the original execution order. Duplicate partitions are removed while preserving the task sequence. The LLM-generated ranking is retained, but compliance with hard constraints is confirmed only through implementation and measurement.

The validated candidates, retrieved evidence, and allocation rationales are presented to the designer for selection. The selected partition is encoded in a machine-readable representation containing the FPGA and GPP task sets, hardware-boundary input/output (I/O), required source files, and implementation configuration. From this fixed representation, the deterministic backend generates the HLS top-level design and testbench, I/O mappings, Vitis HLS and Vivado Tool Command Language (Tcl) scripts, and Python Productivity for Zynq (PYNQ) software. A generation manifest records the selected knowledge-grounded allocation and verifies its source artefacts. The same implementation flow is applied to every validated partition, connecting RAG-assisted LLM reasoning with reproducible hardware generation and on-board evaluation.






\begin{table*}[!t]
\centering
\caption{Gen-TAS rank-1 partitioning, implementation cost, and hardware-cost
reduction across design objectives and LLMs (five runs per configuration).}
\label{tab:llm_partition_results}
\vspace{-2mm}

\begingroup
\fontsize{6.0pt}{7.0pt}\selectfont
\setlength{\tabcolsep}{1.8pt}
\renewcommand{\arraystretch}{1.20}
\setlength{\aboverulesep}{0.30ex}
\setlength{\belowrulesep}{0.30ex}
\setlength{\cmidrulesep}{0.15ex}
\setlength{\cmidrulewidth}{0.3pt}

\resizebox{\textwidth}{!}{%
\begin{tabular}{ccccccccccccccc}
\toprule
\multirow{2}{*}{\textbf{Workload}}
& \multirow{2}{*}{\textbf{Objective}}
& \multirow{2}{*}{\textbf{LLM}}
& \multirow{2}{*}{\textbf{Rank-1}}
& \multirow{2}{*}{\textbf{Consistency}}
& \multirow{2}{*}{\textbf{Latency (ms)}}
& \multirow{2}{*}{\textbf{Speedup}}
& \multicolumn{7}{c}{\textbf{Implementation Metrics}}
& \multirow{2}{*}{\textbf{LLM Time (s)}} \\
\cmidrule(lr){8-14}
& & & & & & &
\textbf{LUT} &
\textbf{FF} &
\textbf{DSP} &
\textbf{BRAM} &
\textbf{FPGA P. (W)} &
\textbf{GPP P. (W)} &
\textbf{Total P. (W)} & \\
\midrule

\multirow{12}{*}{\textbf{CNNImageProc}}

& \multirow{3}{*}{Constrained Latency}
& GPT-5.6 Sol
& FFFFG & 5/5
& $15.494 \pm 0.003$ & $2.445 \pm 0.001\times$
& 6763 (18.7\%)
& 6189 (10.2\%)
& 8 (68.0\%)
& 42 (27.6\%)
& 0.714 (6.9\%)
& 2.738
& 3.451
& $70.293 \pm 5.169$ \\

& & Gemini 3.6 Flash
& FFFFG & 5/5
& $15.494 \pm 0.003$ & $2.445 \pm 0.001\times$
& 6763 (18.7\%)
& 6189 (10.2\%)
& 8 (68.0\%)
& 42 (27.6\%)
& 0.714 (6.9\%)
& 2.738
& 3.451
& $35.994 \pm 8.547$ \\

& & Claude Fable 5
& FFFFG & 3/5
& $15.494 \pm 0.003$ & $2.445 \pm 0.001\times$
& 6763 (18.7\%)
& 6189 (10.2\%)
& 8 (68.0\%)
& 42 (27.6\%)
& 0.714 (6.9\%)
& 2.738
& 3.451
& $85.202 \pm 3.240$ \\

\cmidrule(lr){2-15}

& \multirow{3}{*}{Communication-Aware}
& GPT-5.6 Sol
& FFFFG & 5/5
& $15.494 \pm 0.003$ & $2.445 \pm 0.001\times$
& 6763 (18.7\%)
& 6189 (10.2\%)
& 8 (68.0\%)
& 42 (27.6\%)
& 0.714 (6.9\%)
& 2.738
& 3.451
& $66.908 \pm 4.870$ \\

& & Gemini 3.6 Flash
& FFFFG & 4/5
& $15.494 \pm 0.003$ & $2.445 \pm 0.001\times$
& 6763 (18.7\%)
& 6189 (10.2\%)
& 8 (68.0\%)
& 42 (27.6\%)
& 0.714 (6.9\%)
& 2.738
& 3.451
& $30.467 \pm 0.947$ \\

& & Claude Fable 5
& FFFFG & 5/5
& $15.494 \pm 0.003$ & $2.445 \pm 0.001\times$
& 6763 (18.7\%)
& 6189 (10.2\%)
& 8 (68.0\%)
& 42 (27.6\%)
& 0.714 (6.9\%)
& 2.738
& 3.451
& $86.874 \pm 4.136$ \\

\cmidrule(lr){2-15}

& \multirow{3}{*}{Prioritise Power}
& GPT-5.6 Sol
& GGFFG & 5/5
& $49.202 \pm 0.026$ & $0.770 \pm 0.000\times$
& 4012 (51.7\%)
& 4830 (29.9\%)
& 2 (92.0\%)
& 26 (55.2\%)
& 0.645 (15.9\%)
& 2.737
& 3.383
& $71.699 \pm 8.215$ \\

& & Gemini 3.6 Flash
& FFFFG & 4/5
& $15.494 \pm 0.003$ & $2.445 \pm 0.001\times$
& 6763 (18.7\%)
& 6189 (10.2\%)
& 8 (68.0\%)
& 42 (27.6\%)
& 0.714 (6.9\%)
& 2.738
& 3.451
& $27.253 \pm 1.443$ \\

& & Claude Fable 5
& GGFFG & 5/5
& $49.202 \pm 0.026$ & $0.770 \pm 0.000\times$
& 4012 (51.7\%)
& 4830 (29.9\%)
& 2 (92.0\%)
& 26 (55.2\%)
& 0.645 (15.9\%)
& 2.737
& 3.383
& $98.271 \pm 12.570$ \\

\cmidrule(lr){2-15}

& \multirow{3}{*}{Strict Resource}
& GPT-5.6 Sol
& GGFFG & 3/5
& $49.202 \pm 0.026$ & $0.770 \pm 0.000\times$
& 4012 (51.7\%)
& 4830 (29.9\%)
& 2 (92.0\%)
& 26 (55.2\%)
& 0.645 (15.9\%)
& 2.737
& 3.383
& $67.728 \pm 5.397$ \\

& & Gemini 3.6 Flash
& FFFFG & 5/5
& $15.494 \pm 0.003$ & $2.445 \pm 0.001\times$
& 6763 (18.7\%)
& 6189 (10.2\%)
& 8 (68.0\%)
& 42 (27.6\%)
& 0.714 (6.9\%)
& 2.738
& 3.451
& $34.362 \pm 11.411$ \\

& & Claude Fable 5
& FFFFG & 5/5
& $15.494 \pm 0.003$ & $2.445 \pm 0.001\times$
& 6763 (18.7\%)
& 6189 (10.2\%)
& 8 (68.0\%)
& 42 (27.6\%)
& 0.714 (6.9\%)
& 2.738
& 3.451
& $103.820 \pm 19.449$ \\

\midrule

\multirow{12}{*}{\textbf{CAMC}}

& \multirow{3}{*}{Constrained Latency}
& GPT-5.6 Sol
& GFF & 5/5
& $9.173 \pm 0.052$ & $3.537 \pm 0.020\times$
& 5052 (17.2\%)
& 5930 (13.2\%)
& 21 (27.6\%)
& 32 (3.0\%)
& 0.654 (1.8\%)
& 2.737
& 3.392
& $58.913 \pm 3.129$ \\

& & Gemini 3.6 Flash
& GFF & 5/5
& $9.173 \pm 0.052$ & $3.537 \pm 0.020\times$
& 5052 (17.2\%)
& 5930 (13.2\%)
& 21 (27.6\%)
& 32 (3.0\%)
& 0.654 (1.8\%)
& 2.737
& 3.392
& $22.161 \pm 2.675$ \\

& & Claude Fable 5
& GFF & 4/5
& $9.173 \pm 0.052$ & $3.537 \pm 0.020\times$
& 5052 (17.2\%)
& 5930 (13.2\%)
& 21 (27.6\%)
& 32 (3.0\%)
& 0.654 (1.8\%)
& 2.737
& 3.392
& $83.743 \pm 3.393$ \\

\cmidrule(lr){2-15}

& \multirow{3}{*}{Communication-Aware}
& GPT-5.6 Sol
& GFF & 5/5
& $9.173 \pm 0.052$ & $3.537 \pm 0.020\times$
& 5052 (17.2\%)
& 5930 (13.2\%)
& 21 (27.6\%)
& 32 (3.0\%)
& 0.654 (1.8\%)
& 2.737
& 3.392
& $62.735 \pm 4.117$ \\

& & Gemini 3.6 Flash
& FFF & 4/5
& $0.351 \pm 0.007$ & $92.531 \pm 1.757\times$
& 6102 (0.0\%)
& 6724 (1.6\%)
& 29 (0.0\%)
& 33 (0.0\%)
& 0.666 (0.0\%)
& 2.737
& 3.404
& $23.792 \pm 2.037$ \\

& & Claude Fable 5
& GFF & 5/5
& $9.173 \pm 0.052$ & $3.537 \pm 0.020\times$
& 5052 (17.2\%)
& 5930 (13.2\%)
& 21 (27.6\%)
& 32 (3.0\%)
& 0.654 (1.8\%)
& 2.737
& 3.392
& $76.910 \pm 9.819$ \\

\cmidrule(lr){2-15}

& \multirow{3}{*}{Prioritise Power}
& GPT-5.6 Sol
& GFF & 4/5
& $9.173 \pm 0.052$ & $3.537 \pm 0.020\times$
& 5052 (17.2\%)
& 5930 (13.2\%)
& 21 (27.6\%)
& 32 (3.0\%)
& 0.654 (1.8\%)
& 2.737
& 3.392
& $63.629 \pm 4.788$ \\

& & Gemini 3.6 Flash
& GFF & 5/5
& $9.173 \pm 0.052$ & $3.537 \pm 0.020\times$
& 5052 (17.2\%)
& 5930 (13.2\%)
& 21 (27.6\%)
& 32 (3.0\%)
& 0.654 (1.8\%)
& 2.737
& 3.392
& $24.093 \pm 1.963$ \\

& & Claude Fable 5
& GFF & 4/5
& $9.173 \pm 0.052$ & $3.537 \pm 0.020\times$
& 5052 (17.2\%)
& 5930 (13.2\%)
& 21 (27.6\%)
& 32 (3.0\%)
& 0.654 (1.8\%)
& 2.737
& 3.392
& $86.306 \pm 2.950$ \\

\cmidrule(lr){2-15}

& \multirow{3}{*}{Strict Resource}
& GPT-5.6 Sol
& FGG & 4/5
& $23.553 \pm 0.015$ & $1.378 \pm 0.001\times$
& 4481 (26.6\%)
& 5447 (20.3\%)
& 8 (72.4\%)
& 2 (93.9\%)
& 0.630 (5.4\%)
& 2.737
& 3.367
& $65.645 \pm 6.248$ \\

& & Gemini 3.6 Flash
& GFF & 5/5
& $9.173 \pm 0.052$ & $3.537 \pm 0.020\times$
& 5052 (17.2\%)
& 5930 (13.2\%)
& 21 (27.6\%)
& 32 (3.0\%)
& 0.654 (1.8\%)
& 2.737
& 3.392
& $27.453 \pm 4.208$ \\

& & Claude Fable 5
& GFF & 5/5
& $9.173 \pm 0.052$ & $3.537 \pm 0.020\times$
& 5052 (17.2\%)
& 5930 (13.2\%)
& 21 (27.6\%)
& 32 (3.0\%)
& 0.654 (1.8\%)
& 2.737
& 3.392
& $82.463 \pm 4.900$ \\

\bottomrule
\end{tabular}%
}
\endgroup

\par\vspace{1mm}
\noindent
\begin{minipage}{\textwidth}
\fontsize{6.0pt}{7.0pt}\selectfont
\textit{Note:} F and G denote FPGA and GPP task placement, respectively.
Consistency gives the number of five LLM runs producing the dominant rank-1
partition. Latency and speedup are reported as mean $\pm$ standard deviation
over five hardware deployment runs for each partition. Speedup is relative to
the corresponding all-GPP implementation. For LUT, FF, DSP, and BRAM, values
in parentheses give the percentage reduction relative to the maximum
characterised resource value for the corresponding workload. For FPGA power,
values in parentheses give the percentage reduction relative to the
corresponding all-FPGA implementation (0.767~W for CNNImageProc and 0.666~W
for CAMC). FPGA, GPP, and total power are post-implementation estimates.
LLM time is reported as mean $\pm$ standard deviation over five LLM runs.
\end{minipage}

\vspace{-1.5mm}
\end{table*}

\section{Experimental Evaluation}

\subsection{Experimental Setup}
\label{subsec:experimental_setup}

Experiments are conducted on an AMD ZCU104 platform equipped with an XCZU7EV-2FFVC1156 Zynq UltraScale+ MPSoC \cite{ZCU104,ZynqMPSoC}. Gen-TAS automatically generates the Tcl scripts required by Vitis HLS 2024.1 and Vivado 2024.1 to synthesise the hardware IP, construct and implement the SoC design, produce implementation reports, and generate the bitstream. The automatically generated PYNQ software then controls data transfer, accelerator execution, and on-board evaluation.

The CNNImageProc and Chessboard Automatic Modulation Classification (CAMC) workloads are evaluated under four latency-, communication-, power-, and resource-oriented objectives using OpenAI GPT-5.6 Sol, Google Gemini 3.6 Flash, Anthropic Claude Fable 5. All models receive identical source code, task graphs, requirements, retrieved knowledge, and generation settings. Each workload-objective-LLM configuration is evaluated in five independent sessions.

Gen-TAS is evaluated in terms of allocation validity, requirement satisfaction, consistency, LLM runtime, and implementation performance. A run is successful when it produces a valid allocation and the generated design passes FPGA implementation and on-board functional validation. LUT, FF, DSP, BRAM, maximum frequency, and estimated power are obtained from Vivado post-implementation reports. End-to-end latency, including data transfer, accelerator control, and hardware execution, is measured on the ZCU104 through PYNQ, with outputs verified against the reference software using identical inputs.

\subsection{Case Studies}
\label{subsec:case_studies}

Two representative workloads are used solely to validate Gen-TAS across
different task structures.

\textbf{CNNImageProc:} The Fashion-MNIST CNN processes $1 \times 28 \times 28$ grayscale images and is decomposed into five tasks: \texttt{conv1\_feature\_extract} ($16 \times 28 \times 28$), \texttt{relu\_pool1} ($16 \times 14 \times 14$), \texttt{conv2\_feature\_extract} ($32 \times 14 \times 14$), \texttt{relu\_pool2} ($32 \times 7 \times 7$), and \texttt{dense\_classifier} (10 output classes). These stages expose distinct computational and data-transfer characteristics, making the CNN suitable for fine-grained FPGA-GPP task-allocation evaluation.

\textbf{CAMC:} The SDR case study uses CAMC \cite{CAMC-FPGA}, decomposed into three tasks: \texttt{axis\_initialisation} for SDR I/Q data preprocessing, \texttt{matrix\_generation} for constellation graph-based chessboard-matrix construction, and \texttt{array\_product} for matrix production, template matching and score comparison. These tasks cover control, memory-access, and matrix-operation behaviours, making the design suitable for evaluating FPGA-GPP task allocation.

\subsection{User-Defined Design Objectives}
\label{subsec:objectives}

User-defined objectives are a primary input to Gen-TAS and directly influence the generated FPGA-GPP allocation strategies. The framework is evaluated under four objective directions: constrained latency, communication-aware allocation, power prioritisation, and strict resource utilisation. The detailed requirements and constraint thresholds are defined according to each workload. For CNNImageProc, the all-GPP and all-FPGA latencies are 37.877 and 14.685\,ms, respectively, while the all-FPGA implementation uses 8295 LUTs with a total estimated power of 3.505\,W. For CAMC, the corresponding latencies are 32.445 and 0.356\,ms, with the all-FPGA implementation using 6102 LUTs and a  total estimated power of 3.404\,W. These measurements establish workload-specific design bounds, from which equivalent percentage-based constraints are derived to ensure comparable objective severity across both case studies. Detailed hardware constraints for the CNNImageProc and CAMC designs are available at \url{https://github.com/aprilaihub/Gen-TAS}.

The following examples specify the detailed requirements for each objective:

\textbf{\textit{Constrained Latency:}} Minimise end-to-end latency while targeting at least a 15\% reduction in LUT utilisation and a 0.3\% reduction in total estimated power relative to the all-FPGA baseline, while avoiding communication-dominated partitions. The resulting LUT/ power limits are 7051/3.494\,W for CNNImageProc and 5187/3.394\,W for CAMC. These constraints exclude the all-FPGA endpoint while retaining feasible accelerated partitions in both workloads.

\textbf{\textit{Communication-Aware:}} Minimise end-to-end latency while preferring mixed FPGA-GPP placement when the expected computational benefit exceeds data-transfer, cache-coherency, and control overhead. LUT utilisation and total estimated power should not exceed the corresponding all-FPGA values of 8295 LUTs/3.505\,W for CNNImageProc and 6102 LUTs/3.404\,W for CAMC. This objective tests whether allocation decisions account for hardware-software communication rather than considering acceleration in isolation.

\textbf{\textit{Prioritise Power:}} Minimise total estimated power first and end-to-end latency second, targeting at least a 1\% total power reduction and a 10\% LUT reduction relative to the all-FPGA baseline. The resulting LUT/total power targets are 7466/3.470\,W for CNNImageProc and 5492/3.370\,W for CAMC. This objective tests whether Gen-TAS favours lower-power and lower-resource allocations while preserving useful acceleration.

\textbf{\textit{Strict Resource:}} Apply a restrictive budget requiring at least a 25\% LUT reduction and a 5\% total power reduction relative to the all-FPGA baseline, while minimising latency among mixed partitions. The corresponding limits are 6221 LUTs/3.330\,W for CNNImageProc and 4577 LUTs/3.234\,W for CAMC. These limits are intentionally outside the characterised feasible power region and therefore serve as an infeasibility stress test. When no candidate satisfies both limits, Gen-TAS is instructed to return the three least violating candidates rather than report the constraints as satisfied. 

The resulting allocations are evaluated against these objectives in Table~\ref{tab:llm_partition_results}.

\subsection{Quantitative Results}

Table~\ref{tab:llm_partition_results} demonstrates that Gen-TAS generates requirement-driven allocations rather than applying a fixed partition across objectives. For CNNImageProc, constrained-latency and communication-aware objectives consistently favour \texttt{FFFFG}, achieving a 2.45$\times$ speedup while reducing LUT, FF, DSP, and BRAM usage by 18.7\%, 10.2\%, 68.0\%, and 27.6\%, respectively. Under the power-prioritisation objective, GPT and Claude select \texttt{GGFFG}, reducing FPGA power by 15.9\% and LUT, FF, DSP, and BRAM usage by 51.7\%, 29.9\%, 92.0\%, and 55.2\%, respectively, at the cost of increased latency. These results illustrate the workload-specific trade-off between acceleration and implementation cost.

For CAMC, \texttt{GFF} is selected for most objectives, providing a 3.53$\times$ speedup with reductions of 17.2\% LUTs, 13.2\% FFs, 27.6\% DSPs, and 1.8\% FPGA power. Under the communication-aware objective, Gemini selects the all-FPGA \texttt{FFF} partition, achieving the highest speedup of 92.53$\times$. Conversely, GPT selects \texttt{FGG} for the strict-resource objective, reducing LUT, FF, DSP, and BRAM usage by 26.6\%, 20.3\%, 72.4\%, and 93.9\%, respectively, while retaining a 1.38$\times$ speedup and reducing FPGA power by 5.4\%. This confirms that Gen-TAS can adapt task allocation to latency, communication, power, and resource requirements, although the selected trade-offs vary across LLMs.

\begin{figure}[!t]
    \centering

    \includegraphics[
        width=\columnwidth,
        trim={0cm -1cm 0cm 0cm},
        clip
    ]{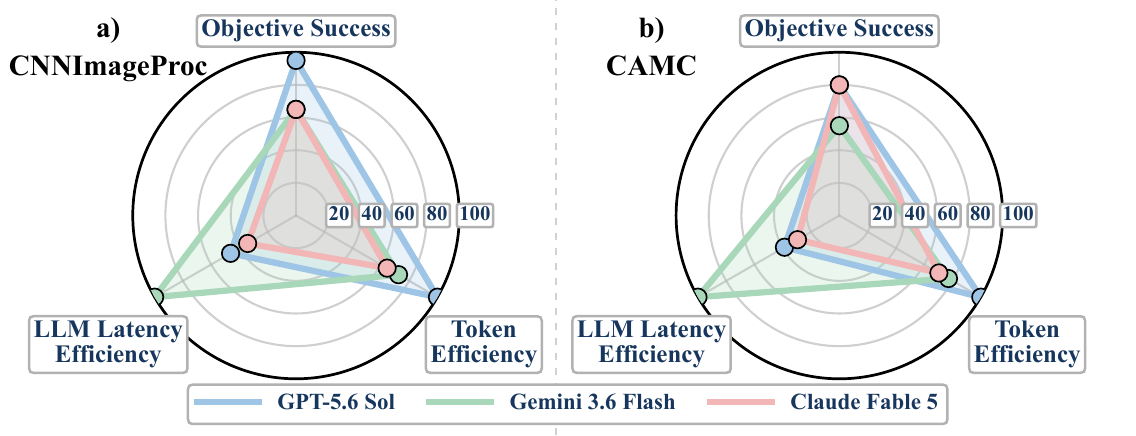}
    
    \caption{LLM comparison for (a) CNNImageProc and (b) CAMC across
    objective satisfaction, token efficiency, and LLM latency efficiency.
    Results aggregate five independent runs across four Gen-TAS design objectives for each LLM. Higher values indicate better performance.}
    \label{fig:radar}
    \vspace{-2mm}
\end{figure}

The repeated experiments also demonstrate stable allocation generation under the RAG-grounded prompting flow. Of the 24 workload-objective-LLM configurations, 22 reproduce the dominant partition in at least four of five runs, including 15 with complete five-of-five consistency. Fig.~\ref{fig:radar} complements the implementation results by comparing objective satisfaction, token efficiency, and LLM latency efficiency. Gemini generally achieves the highest latency efficiency, with mean generation times of 22.161-35.994s, compared with 58.913-71.699s for GPT and 76.910-103.820s for Claude. Overall, the results demonstrate the model-agnostic capability of Gen-TAS to generate stable, reproducible, and user-requirement-driven FPGA-GPP allocations across different LLMs.


\section{Conclusion}

This work proposed Gen-TAS, an LLM-assisted framework that integrates RAG and implementation knowledge for requirement-driven task allocation in FPGA-GPP heterogeneous systems. Evaluations across multiple LLMs, user-defined objectives, and application domains demonstrate successful, stable, and reproducible generation of practical hardware-software allocations. Gen-TAS therefore bridges natural-language design requirements and executable heterogeneous implementations while reducing manual exploration and the expertise barrier to FPGA-GPP design.

Future work will expand the knowledge base and incorporate implementation feedback to improve retrieval quality and allocation robustness. Further directions include finer-grained partitioning, multi-accelerator generation, more heterogeneous platforms, and enhanced LLM prompting and retrieval.

\section*{Acknowledgment}
The authors thank the AMD University Programme for Vitis/Vivado licenses and FPGA boards donation.







\end{document}